\documentclass[11pt]{article}
\usepackage{amsmath, amsthm, amssymb,color,fullpage,url,booktabs}  
\usepackage{graphicx}

\usepackage{cleveref}
\usepackage{tikz}

\newcommand{\nc}{\newcommand}
\nc{\rnc}{\renewcommand}

\newcommand{\ket}[1]{\left|#1\right\rangle}
\newcommand{\proj}[1]{\left|#1\right\rangle\left\langle #1\right|}

\DeclareMathOperator{\poly}{poly}

\DeclareMathOperator{\supp}{supp}

\def\be#1\ee{\begin{equation}#1\end{equation}}
\def\ba#1\ea{\begin{align}#1\end{align}}
\def\bas#1\eas{\begin{align*}#1\end{align*}}
\def\bpm#1\epm{\begin{pmatrix}#1\end{pmatrix}}
\nc{\non}{\nonumber}
\nc{\nn}{\nonumber}
\nc{\eq}[1]{(\ref{eq:#1})}
\nc{\eqs}[2]{(\ref{eq:#1}) and (\ref{eq:#2})}
\rnc{\L}{\left} 
\nc{\R}{\right}
\nc{\ra}{\rightarrow}
\nc{\ot}{\otimes}
\nc{\grad}{{\vec{\nabla}}}

\def\bea#1\eea{\begin{eqnarray}#1\end{eqnarray}}
\def\beas#1\eeas{\begin{eqnarray*}#1\end{eqnarray*}}

\newtheorem{thm}{Theorem}
\newtheorem*{thm*}{Theorem}

\newtheorem{proto}{Protocol}

\theoremstyle{definition}

\newtheorem{dfn}[thm]{Definition}
\theoremstyle{plain}

\makeatletter
\newtheorem*{rep@theorem}{\rep@title}
\newcommand{\newreptheorem}[2]{%
\newenvironment{rep#1}[1]{%
 \def\rep@title{#2 \ref{##1} (restatement)}%
 \begin{rep@theorem}}%
 {\end{rep@theorem}}}
\makeatother

\newreptheorem{thm}{Theorem}
\newreptheorem{lem}{Lemma}

\nc\eps{\epsilon}

\nc\cA{\mathcal{A}}
\nc\cB{\mathcal{B}}
\nc\cC{\mathcal{C}}
\nc\cD{\mathcal{D}}
\nc\cE{\mathcal{E}}
\nc\cF{\mathcal{F}}
\nc\cG{\mathcal{G}}
\nc\cH{\mathcal{H}}
\nc\cI{\mathcal{I}}
\nc\cJ{\mathcal{J}}
\nc\cK{\mathcal{K}}
\nc\cL{\mathcal{L}}
\nc\cM{\mathcal{M}}
\nc\cN{\mathcal{N}}
\nc\cO{\mathcal{O}}
\nc\cP{\mathcal{P}}
\nc\cQ{\mathcal{Q}}
\nc\cR{\mathcal{R}}
\nc\cS{\mathcal{S}}
\nc\cT{\mathcal{T}}
\nc\cU{\mathcal{U}}
\nc\cV{\mathcal{V}}
\nc\cW{\mathcal{W}}
\nc\cX{\mathcal{X}}
\nc\cY{\mathcal{Y}}
\nc\cZ{\mathcal{Z}}

\nc\bbC{\mathbb{C}}

\nc\bbF{\mathbb{F}}
\nc\bbM{\mathbb{M}}
\nc\bbN{\mathbb{N}}
\nc\bbR{\mathbb{R}}
\nc\bbZ{\mathbb{Z}}

\nc\benum{\begin{enumerate}}
\nc\eenum{\end{enumerate}}
\nc\bit{\begin{itemize}}
\nc\eit{\end{itemize}}

\newcommand{\secref}[1]{Section~\ref{sec:#1}}

\newcommand{\tabref}[1]{Table~\ref{tab:#1}}

\nc{\todo}[1]{\textcolor{red}{todo: #1}}
\nc{\Anote}[1]{\textcolor{red}{Aram note: #1}}

\def\begsub#1#2\endsub{\begin{subequations}\label{eq:#1}\begin{align}#2\end{align}\end{subequations}}
\nc\qand{\qquad\text{and}\qquad}
\nc\mnb[1]{\medskip\noindent{\bf #1}}

\nc{\pder}[2]{\frac{\partial {#1}}{\partial {#2}}}
\nc{\p}{\partial}
\usepackage{algorithm, algorithmicx, float, comment,tabularx}
\usepackage{etoolbox}
\usepackage{algpseudocode}

\newtoggle{long}
\nc{\longonly}[1]{\iftoggle{long}{#1}{}}
\nc{\shortonly}[1]{\iftoggle{long}{}{#1}}
\nc{\longshort}[2]{\iftoggle{long}{#1}{#2}}
\toggletrue{long}

\nc{\boxwidth}{\longshort{0.96}{0.48}\textwidth}

\usepackage{makecell}

\usepackage{cellspace}
\begin{document}

\title{Small quantum computers and large classical data sets }
\longshort{\author{Aram W. Harrow\footnote{Center for Theoretical Physics,
      Massachusetts Institute of Technology, {\tt aram@mit.edu}}}}
{\author[a]{Aram W. Harrow}
\affil[a]{Center for Theoretical Physics, Massachusetts Institute of Technology, Cambridge, MA 02139}}


\longonly{\maketitle }
\begin{abstract}
  We introduce hybrid classical-quantum algorithms for problems involving a
  large classical data set $X$ and a space of models $Y$ such that a quantum
  computer has superposition access to $Y$ but not $X$.  These algorithms use
  data reduction techniques to construct a weighted subset of $X$ called a
  coreset that yields approximately the same loss for each model.  The coreset
  can be constructed by the classical computer alone, or via an interactive
  protocol in which the outputs of the quantum computer are used to help decide
  which elements of $X$ to use.  By using the quantum computer to perform Grover
  search or rejection sampling, this yields quantum speedups for maximum
  likelihood estimation, Bayesian inference and saddle-point
  optimization. Concrete applications include k-means clustering, logistical
  regression, zero-sum games and boosting.
\end{abstract}

\shortonly{\maketitle }

\section{Introduction}\label{sec:overview}


What can a quantum computer do with a large classical data set?  At first glance it would
seem that the costs of loading the data into the quantum computer would overwhelm any
possible quantum speedup.   And how can Grover's algorithm be used for practical problems?
While the original title of Grover's paper~\cite{Grover96} is ``A fast quantum mechanical algorithm for
database search,'' its applications so far have mostly been to exploring combinatorial
search spaces.  Similar questions apply to other quantum optimization algorithms, such as
the adiabatic algorithm~\cite{farhi00}, quantum walks~\cite{krovi15}, and the QAOA~\cite{QAOA}.  

Existing quantum algorithms for optimization and machine learning are often less complete
than their classical counterparts because they do not use realistic models of their input
data~\cite{aaronson2015}.  One way to view this is that they are meant to be subroutines in larger
``end-to-end'' algorithms that will provide the data in the needed format.  However, there
has been relatively little research on these more complete algorithms and their
development has often been nontrivial.  Indeed, the trivial methods of turning large
classical datasets into either quantum oracles or quantum states are so expensive as to
negate any possible quantum advantage.  As a result, even Grover's algorithm has not yet
been successfully applied to speed up any natural machine learning task, despite being
arguably the simplest, most widely known, and most widely applicable quantum algorithm.

One proposal is to use a ``quantum RAM,'' typically meaning a large
classical memory which can be queried in superposition~\cite{GLM07,GLM08}.  This enables
powerful quantum algorithms to be used, often with provable speedups.  However, current
classical computing or data storage hardware does not function as quantum RAM, and
near-term hardware plans by leading groups using trapped ions, superconductors, or photons
on chips do not involve quantum RAM.   Also, building a large quantum
RAM may run into many of the same challenges that occur in building a large universal
quantum computer~\cite{Aruna15}.

At first glance it seems that quantum advantage requires problems with small input sizes.
If the input size is $n$ then it is hard to avoid expending effort proportional to $n$
even before the computation starts, in order to acquire, store and load the input.  Even
alternate models such as property testing or streaming can typically mitigate only some of
these.  Similar issues apply to large outputs.  For problems that can be solved on
classical computers in time $\tilde O(n)$, it would seem that there is little scope for
quantum advantage.  For this reason, proposals for quantum advantage usually involve
problems where the best known classical runtime scales rapidly with the input size,
perhaps exponentially.

At the same time, many tasks in classical computing, such as machine learning, are
evolving towards the use of increasingly large data sets, for which a runtime of even
$O(n^2)$ can be infeasible. How can quantum computers be of use in this setting?

This paper will explore the ability of small quantum computers to work together with large
classical computers to analyze large data sets.  We will work in a regime where our input
size is $n$, the classical computer runs in time nearly linear in $n$, and we do not use
any unconventional access models for the quantum computer.  Since
qubits are likely to always remain more expensive than bits, this hybrid classical-quantum
model should be relevant even when quantum computers with millions or billions of qubits
are available.

The key principle will be {\em data reduction}.  We will approximate a data set
$X = \{x_1,\ldots,x_n\}$ with a much smaller subset $X'$ (called a ``coreset'')
along with a weight function $w: X' \ra \bbR_{\geq 0}$ such that $X',w$ can
adequately substitute for $X$ in solving problems of interest.  This can be
achieved either by a classical computer alone, or a classical computer together
with a quantum computer.  Then a hard optimization problem can be solved on the
quantum computer using the reduced data set $X'$.

Data reduction can be useful for any quantum optimization algorithm for which
computing the objective function requires examining a large data set.  We
illustrate its benefits with a representative example in
\secref{motivating-example} before giving an informal overview of our algorithms
and their benefits in \secref{data-reduction} and discussing input data models
in \secref{data-models}.  The rest of the paper spells out the results more
formally, beginning with formal descriptions of the problem setting in
\secref{setting} and the previously developed algorithms in \secref{known}, and
proceeding to describe our algorithms in detail in
\secref{algorithms}.  There are several relevant related algorithms,
such as variational quantum algorithms and stochastic gradient
descent, and we compare our approach with these in \secref{compare}
before concluding in \secref{concl}.

\subsection{Motivating example: minimizing empirical loss with Grover.}
\label{sec:motivating-example}
Suppose we are
given a set of data points $X$, a set of models $Y$, and a loss function
$f:X\times Y\mapsto \bbR$.  Our goal is to compute \be \arg\min_{y\in Y} \sum_{x\in X}
f(x,y),\label{eq:min-sum}\ee i.e.~to choose the model which minimizes the empirical loss.
It is important to emphasize the form of the input: $X$ is an explicit data set
$x_1,\ldots,x_n$; $Y$ is a set that may be large or infinite but has a succinct
description, e.g.~$Y$ might be the set of all ways of choosing a mixture of up to $k$
Gaussians in $\bbR^d$; $f$ is given as an explicit and short algorithm.

To apply Grover's algorithm
(technically the D\"urr-H\o{}yer algorithm for minimizing a black-box
function~\cite{durr96}) will require $O(\sqrt{|Y|})$ evaluations of
$F(y) := \sum_{x\in X}f(x,y)$.  However, each evaluation of $F$ requires iterating over
the entire data set $X$.  This takes time $O(|X|)$ if we assume for simplicity that $f$ can be
computed in time $O(1)$.  The crucial feature of this problem is that the set $Y$ can be
accessed in superposition, so that we can obtain the quadratic Grover speedup in searching
over it, but $X$ is a classical data set which cannot be queried in this way.  Thus we
could not use quantum algorithms such those for approximate counting~\cite{BHMT02} to
speed up the evaluation of $F(y)$.  As a result, the classical runtime
of $O(|X|\cdot |Y|)$ turns into a quantum runtime of $O(|X|\cdot
\sqrt{|Y|})$.  If $|X|$ is comparable to $|Y|$, then this erodes much
of the savings from Grover's algorithm.

Instead we will use a coreset $X'$ (with weight function $w$) and replace $F(y)$
with its approximation $F_w(y) := \sum_{x \in X'} w(x) f(x,y)$.  This results in a
hybrid classical-quantum algorithm for the overall problem.  A classical
computer needs to examine the original data set $X$ in order to calculate $X'$
and then a quantum computer can minimize $F_w$ in time $O(\sqrt{|Y|}|X'|)$.  If
$|X'|\ll |X|$ then this yields a nearly quadratic speedup, and if
$\min_y F_w(y) \approx \min_y F(y)$ then this provides a good approximation to
the original problem.  Coresets satisfying both of these properties are known in
a large number of cases, as we discuss below in
\Cref{sec:coreset-review,sec:algorithms}.

In some cases (see \Cref{sec:algorithms}) the size of $X'$ will depend only on the level of
approximation desired and not on the size of the original data set
$X$.  When this happens, the classical or quantum runtime will not
depend on the product of $|X|$ and $|Y|^{1 \text{ or }\frac 12}$, but
instead on (roughly) their sum.  Coresets speed up both classical and
quantum algorithms, but they increase the relative quantum speedup by
reducing the time spent on tasks where there is no known quantum advantage.

\subsection{More general uses of coresets}\label{sec:data-reduction}
There are three main directions in which this basic example can be modified.  
\bit
\item The D\"urr-H\o{}yer minimization algorithm could be replaced by any other quantum
  algorithm for minimizing functions, such as adiabatic optimization or QAOA.  In almost
  any such algorithm, either $F(y)$ or its gradients will need to be evaluated,
  and the cost of doing so will scale linearly with $|X|$ (see \Cref{sec:known-opt} for
  details). Thus, using a coreset can provide significant savings.  We explore these more
  in Algorithms 1 and 1.1 in \Cref{sec:algorithms}.
  
\item The form of \eq{min-sum} could be substantially changed.  A small change would be to
  minimize $F(y) := r(y)+\sum_x f(x,y)$, where $r(y)$ is a regularizer, perhaps intended to favor
  simpler models.  A bigger change would be to perform Bayesian inference.  As we will
  discuss below, Bayesian inference can be described as
  sampling from a distribution $\pi(y) \propto \exp(F(y))$.  Here too quantum algorithms can
  achieve roughly quadratic speedups provably, and heuristic algorithms have been proposed which
  may have better performance (see \cref{sec:sampling-review}). In each case, quantum speedups are not known for iterating
  over $X$, and so reducing the size of $X$ would increase the relative quantum speedup.  One benefit of sampling over optimization is that the samples output by the
  quantum computer could be used by the classical computer to adaptively augment the
  coreset. This idea is explored in Algorithms 2 and 2.1 of \Cref{sec:algorithms}.
  
\item The coreset can be built iteratively using interaction between the classical and
  quantum processors.  First the classical computer produces a coreset $X'_1$ which the
  quantum computer uses to produce output $y_1$.  Then $y_1$ is used by the classical
  computer to produce a new coreset $X_2'$, which the quantum computer uses to product
  output $y_2$, and so on for $r$ rounds. The final answer could either be $y_r$ or some
  average of $y_1,\ldots,y_r$.  See Algorithm 3 in \secref{algorithms} for details.
\eit

The common theme in these algorithms is that there is an outer loop
involving $Y$ and an inner loop involving $X$.  This outer loop could
involve iterating over all elements of $Y$, performing a Grover-style
search, using adiabatic optimization, or other classical or quantum
algorithms.  Suppose that in general this outer loop requires
$\tau_{\text{outer}}(Y)$ iterations (e.g. $|Y|$ for classical brute-force search,
$\sqrt{|Y|}$ for Grover, and so on).  Then if the inner loop sums over all of $X$,
the overall algorithm will require $O(|X|\tau_{\text{outer}}(Y))$  evaluations of
$f$.  Suppose we have a coreset $X'$ that can be constructed in
time $\tau_{\text{core}}(X)$.  Then our total (classical + quantum) run-time becomes
\be O(\underbrace{\tau_{\text{core}}(X)}_{\text{classical}} +
\underbrace{|X'|\tau_{\text{outer}}(Y)}_{\text{classical or quantum}}).\ee    Assume that
$\tau_{\text{core}}(X)$ scales roughly with $|X|$ while $|X'|$ is roughly independent of
$|X|$ and is determined instead by the complexity of the model and the desired
accuracy. 
Then we have again replaced a run-time that scales as the product of $|X|$ and
$\tau_{\text{outer}}(Y)$ with one that scales roughly as their sum.  If the run-time is
dominated by the complexity of searching over $Y$ then this will increase the relative
quantum speedup.

We illustrate this point with a plot.  Suppose for
simplicity that $|X|=n$, $|X'|=O(1)$,
$\tau_{\text{core}}(X)=n^\alpha$, and $\tau_{\text{outer}}^{\text{classical}}(Y) =
n^\beta$, for some
constants  $\alpha,\beta > 0$. Assume as well that the quantum speedup
is quadratic (e.g.~based on Grover) so that
$\tau_{\text{outer}}^{\text{quantum}}(Y)=n^{\beta/2}$.  We can
summarize the effects of both coresets and classical-vs-quantum
computing as follows.

\begin{center}
\begin{tabular}{ccc}
\toprule
Using coreset? & No & Yes \\ \midrule
Classical & $n^{1+\beta}$ & $n^\alpha + O(n^\beta)$ \\
Quantum & $n^{1+\beta/2}$ & 
$n^\alpha$ classical and $O(n^{\beta/2})$ quantum\\
\bottomrule
\end{tabular}
\end{center}

In the lower-right box, the algorithm uses both classical and quantum
resources, since the classical computer constructs the coreset in time
$n^\alpha$ and the quantum computer performs the optimization in time
$O(n^{\beta/2})$.  The situation is similar if we replace a simple
coreset with an iterative construction.

\begin{figure}[h!]
{\centering 
\includegraphics[width=0.5\textwidth]{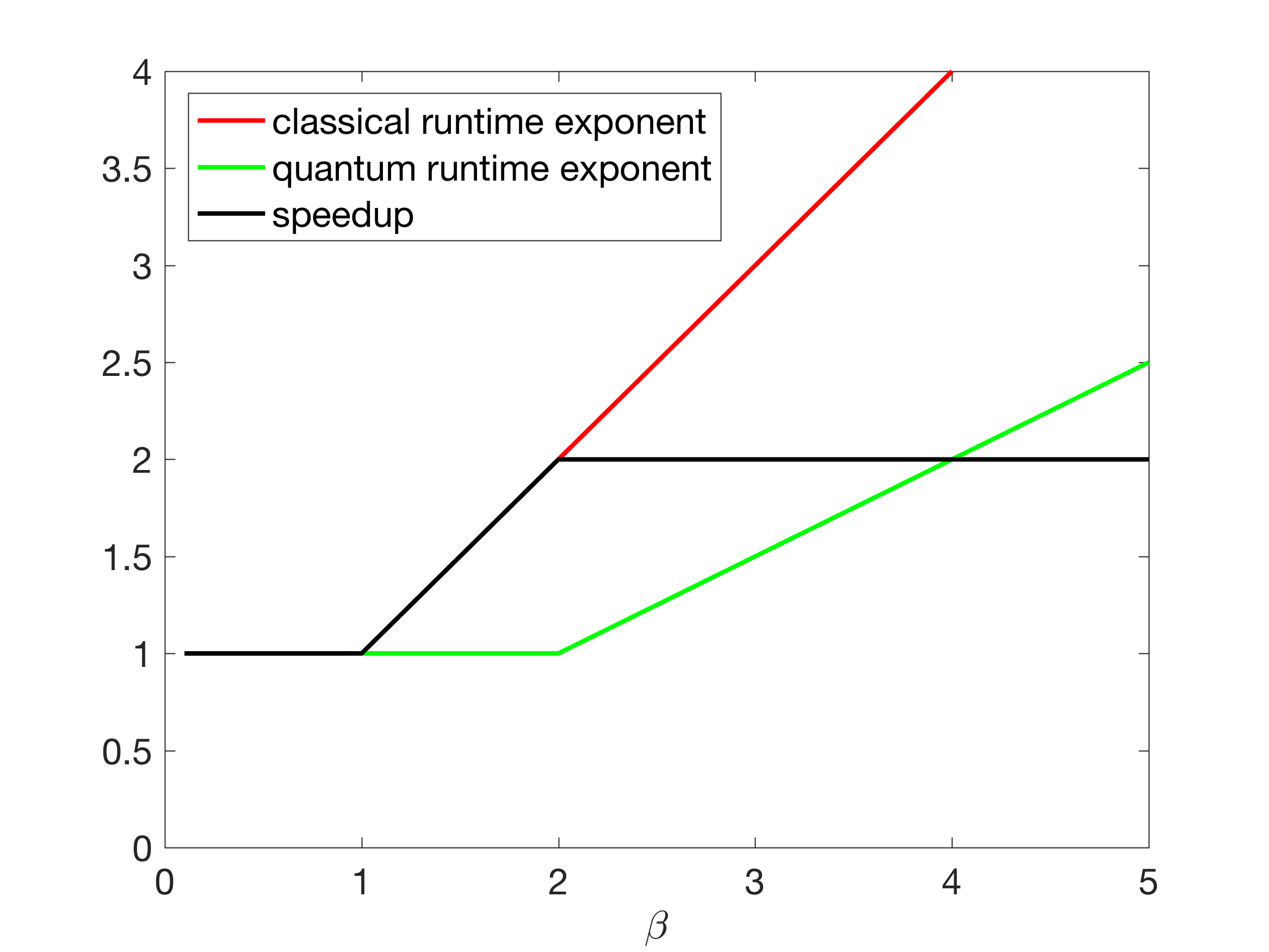}
\caption{Classical and quantum runtime exponents, as well as speedup, as a function of
  $\beta$.  Assume that $\alpha=1$.}\label{fig:speedup}}
\end{figure}

To define the speedup, we say that if the classical and quantum runtimes are
$T_{\text{cl}},T_{\text{q}}$ respectively then the speedup is
$\log(T_{\text{cl}})/\log(T_{\text{q}})$.  Thus ``1'' means no speedup, ``2'' is the
Grover speedup, $\infty$ would mean a superpolynomial (e.g.~exponential) speedup and $<1$
would mean a slowdown.  Suppose that $\alpha=1$, since this is often achievable (as we
discuss below).   The resulting quantum speedups as a function of $\beta$ are illustrated
in Fig.~\ref{fig:speedup}.

\subsection{Data model}\label{sec:data-models}
An important departure of this work from much of the quantum
machine learning literature is in the choice of input data model. This seemingly mundane
issue turns out to be crucial to understanding the utility of many quantum algorithms and
in this section we will broadly review the role of different data models in quantum algorithms.

Existing quantum algorithms have used several different input models.  The
Standard model is a string of bits $x = (x_1,\ldots,x_n)$ which could be
thought of as either as labeling a standard basis state $\ket x$ that is an
input to a quantum circuit, or as input to a classical computer which generates
a quantum circuit $C_x$ which is applied to a fixed input.  The former
interpretation is the standard theoretical model of quantum circuits, while the
latter is closer to how quantum computers would be likely to function in
practice.  Another important model is the Oracle model in which the quantum
computer is given access to a unitary $O$ such that
$O\ket{i,a} = \ket{i,a\oplus x_i}$.  This would arise most naturally if we are
given a (classical or quantum) circuit that can compute $x_i$ given input $i$.
Other uniquely quantum models exist as well.  In the Quantum Data
model, the input is given as an arbitrary $n$-qubit quantum state $\ket\psi$.
Another model is the Quantum Oracle, meaning black-box access to an
$n$-qubit unitary $U$.

All of these models have been widely used in the quantum algorithms literature.
Prominent examples of algorithms using each model are summarized in
\tabref{input-models}.

\begin{table*}[h]
\renewcommand{\arraystretch}{2} 
\begin{tabularx}{\linewidth}{l l X}
\hline
\thead{Input Model} & \thead{Definition}              & \thead{Examples}  \\ \hline
Standard & $x=(x_1,\ldots,x_n) \in \{0,1\}^n$ & Factoring and other number theory
problems. \newline 3-SAT and combinatorial optimization.
\newline variational quantum eigensolver \\
Oracle & $O\ket{i,a} = \ket{i,a+x_i}$ & OR (Grover search), max, approximate counting \newline
NAND tree, collision, [graph] property testing \newline
hidden subgroup problem, welded trees \\
Quantum Data & given state $\ket \psi$ & quantum Fourier transform \newline
 SWAP test, Schur transform, state tomography\newline
Hamiltonian simulation, linear systems solver
\newline learning with quantum examples \\
Quantum Oracle & 
\makecell[t]{given access to unitary $U$\\ and controlled unitary $C_U$}
& phase estimation. \newline quantum sensing and process tomography \newline
qubitization and singular value transform 
\end{tabularx}

\caption{Examples of algorithms using each input model.  In the Quantum Data
  model we sometimes assume instead the ability to perform $V$ and $V^\dag$ for
  some unitary satisfying $V\ket 0 = \ket\psi$.  In the Quantum Oracle model the
  controlled unitary $C_U$ is of the form $\sum_{t=0}^{T-1} \proj t \ot U^t$.
  Here we can take $T=2$ in some cases or can take $T$ to be exponentially large
  in other cases; see \cite{Atia17} for discussion.  }
\label{tab:input-models}
\end{table*}

Many papers on quantum machine learning~\cite{biamonte2016} use the
Oracle (or quantum
RAM~\cite{GLM07,GLM08}) model or the Quantum Data model.
These models are closely related because states of the form $2^{-n/2}\sum_{x\in \{0,1\}^n} e^{i\phi_x}\ket{x}$ can be
prepared easily using an oracle for $\phi_x$ and even general states $\sum_x \alpha_x
\ket{x}$ can be prepared fairly efficiently using an oracle for $\alpha_x$ as long as
$2^{-n/2}\sum_x |\alpha_x|$ is not far from 1.
Current experimental plans
for quantum computing do not correspond to this model and some
discussions of the difficulties of quantum RAM can be found in
\cite{Aruna15,aaronson2015,ciliberto2017}.  Moreover, even under
optimistic assumptions about quantum RAM, a data set of size $N$ will
have cost scaling at least with $N$: both a ``classical'' cost in
acquiring and storing the data, and ``quantum'' cost in pieces of
hardware that can interact with qubits without decohering them.  But
once we have $O(N)$ pieces of quantum hardware, we might use other
algorithmic approaches, such as parallelism; see \cite{GR-memory-04}
for a discussion in a related context.

Some algorithms use multiple data models.  For example, Hamiltonian simulation
and the linear systems solver may use oracles to specify entries of the
matrices, or these could be specified explicitly as sums of Paulis or
creation/annihilation operators.  The recent quantum LP and SDP
algorithms~\cite{BrandaoS16,1710.02581,1809.00643,1809.01731,vAG18} use a
combination of oracles and quantum data, and \cite{vAG18} also has an algorithm
for the Quantum Oracle model.

The input model in this paper can be thought of as a hybrid between the standard and oracle
models.  We are given inputs $X = \{x_1,\ldots,x_n\}$ explicitly, say as records on a hard
drive.  Here the $x_i$ are typically not bits but belong to some alphabet $\cX$.
We are also given oracle access to a function $f:\cX\times Y \mapsto
\bbR$.  Here  $Y$ is either the set $[m] := \{1,2,\ldots,m\}$, or some
other set that can be bijectively and efficiently mapped to $[m]$.
For this model to be realistic,  $f$ should be easy to compute.

A concrete example of this model is $k$-means clustering in some
metric space $\cX$ with distance function $\text{dist}(\cdot,\cdot)$, e.g.~$\bbR^d$ with Euclidean norm.
We are given points $x_1,\ldots,x_n\in \cX$ and are looking for
cluster centers $y_1,\ldots,y_k\in \cX$.  Given $x\in \cX$ and $y=(y_1,\ldots,y_k)\in(\bbR^k)^{\times d}$ the loss is
\be f(x,y) = \min_{i\in [k]} \text{dist}(x,y_i)^\rho.\ee
Here $\rho=2$ for $k$-means clustering, $\rho=1$ for $k$-median
clustering, and other values of $\rho>0$ can be chosen to tune the relative importance of
outliers.
If $\text{dist}$ is easy to compute (e.g. the Euclidean norm) then $f$
is easy to compute as well, and so the oracle assumption is
reasonable.  The set $Y$ here is just the set of points $\cX$, 
made finite by considering only points on a discrete grid within some
bounding box.  Such a simple set is easy to map bijectively to $[m]$
for some large integer $m$.  Finally, the most natural way to expect
the input $x_1,\ldots,x_n$ for a clustering problem is as data on a
classical storage medium.

The hybrid model in this paper works naturally with classical data stored in the same form
that would be used for classical algorithms.  There is only a minor way in which current
quantum computing hardware does not meet its requirements, which is that pulse generators
are often slow to reprogram.  This makes it easy to repeatedly run the
same quantum algorithm but more expensive to modify the gate set
either between iterations or during a single run of the quantum
computer. However, this latency will have to be improved much more to
meet the requirements of fault-tolerant quantum computing.  It also appears to be a
limitation that is not fundamental but rather applies to current off-the-shelf technology
which was originally designed for other tasks.

This paper is not the only one with optimization algorithms that run
on current models of hybrid classical/quantum computers.  There has
been a recent explosion of interest in variational quantum algorithms,
such as those for ground states of Hamiltonians~\cite{Peruzzo13},
constraint satisfaction problems~\cite{QAOA,QAOA2},
learning~\cite{farhi2018classification,mitarai2018quantum,schuld2018circuit}
and other tasks.  These algorithms have the advantage of running on
near-term hardware~\cite{FGGN17} with modest requirements for gates,
connectivity and qubits.  They also make use of a classical computer
to run an algorithm such as gradient descent in the outer loop while
the quantum computer is used to evaluate the cost function or its
gradients in the inner loop.  In this way, variational algorithms make
use of the longer memory lifetime of the classical computer.  By
contrast, our algorithms also use the larger storage of the classical
computer.  Depending on the computational cost of constructing the
coreset, our algorithms may also use many more gates from the
classical computer.  We return to this comparison in
\Cref{sec:compare-VQE} after describing our algorithms.

\section{Problem setting}\label{sec:setting}

The algorithms introduced in this paper involve a choice of  computational tasks, of
quantum optimization or sampling subroutines and of classical data reduction subroutines.
In this section we formally describe these computational tasks and review the quantum and
classical subroutines that we will make use of.

\paragraph{Common assumptions and notation.}
In each task, we are given sets $X,Y$ with $|X|=n$ and $|Y|=m$, along
with a function $f:X\times Y \mapsto\bbR$, or a function $f:\cX\times Y \mapsto \bbR$ for
some $\cX \supseteq X$.  The set $X =
\{x_1,\ldots,x_n\}$ is stored in classical memory (say a hard drive)
while the elements of $Y$ can be implicitly described.   In other
words, there is a bijection $\varphi:[m]\mapsto Y$ such that
$\varphi,\varphi^{-1}$ are both efficiently computable.  For
simplicity, we could also assume that $Y= [m]$.  We assume that a classical algorithm for
computing $f$ is known, and our results will be most relevant when its runtime is small.

\paragraph{Tasks}
This paper focuses on three tasks arising in data analysis.  
\paragraph{Maximum a posteriori estimation (MAP)}
 Suppose further we are given an easily
  computable function $r:Y\mapsto \bbR$.  Define the log-likelihood function
\be F(y) :=
r(y) +\sum_{x \in X} f(x,y)
\label{eq:F-def}\ee
The goal  is to compute 
\be \arg\max_{y\in Y} F(y)\label{eq:MLE}\ee

To connect this rather general optimization question to MLE, we make the following interpretations.
Define $p(x|y) := \exp(f(x,y)) / Z_x$ to be the probability of observing data point $x$ given a
 model $y$, and $\pi_{\text{prior}}(y) := \exp(r(y))/Z_0$ to be the prior distribution.  Here $\{Z_x\}_{x\in X}, Z_0$ are
 normalization factors that do not need to be known to the algorithm.  (Note
 that this interpretation, especially with nonzero $r(y)$, is typically referred
 to as MAP, or maximum a posteriori estimation.)

Technically we do not need an underlying probabilistic model.  If $f(x,y)$ is the loss
that model $y$ incurs from data point $x$, then \eq{MLE} corresponds to {\em empirical
loss minimization}.  An example is clustering, where the loss is a function only of the
distance to the nearest cluster center.

  \paragraph{Bayesian inference}
  Given a prior distribution $\pi_0(y)$, define the posterior distribution 
\ba \pi(y) & = \pi_0(y)
\frac{\exp\left(\sum_{x\in X} f(x,y)\right)}
{\sum_{y'\in Y}\pi_0(y')
  \exp\left(\sum_{x\in X}f(x,y')\right)}
\shortonly{\nn \\ &}
=
\frac{\exp(F(y))}{Z}.
\ea
The second equality uses the definitions of $F,r,Z$ from the MLE task.
The goal here is to sample from $\pi_{\text{posterior}}(y)$.

 \paragraph{Saddle-point optimization.}
The above two problems can be thought of as
taking the max of a sum and sampling from a distribution defined by a sum.  A third task
is to take the max of a min (or a min of a max), a problem sometimes known as saddle-point
optimization.  We describe a version of the problem where the max and min are taken
over probability distributions.

Let $\Delta(S)$ denote the set of probability distributions over a set $S$.
A typical saddle-point optimization problem is to compute
\be \max_{\theta_Y\in \Delta(Y)} \min_{x\in X} \sum_y \theta_y f(x,y)
 = \min_{\theta_X\in \Delta(X)} \max_{y\in Y} \sum_x \theta_x f(x,y)
 \label{eq:zero-sum}\ee
The equality here is due to von Neumann's minimax theorem.  We could equivalently write
$\max_{\theta_Y\in \Delta(Y)}\min_{\theta_X \in\Delta(X)}$ or
$\min_{\theta_X \in\Delta(X)}\max_{\theta_Y\in \Delta(Y)}$.
As in the previous problems, $X$ is a set of records in a classical data set and $Y$ is a
set that can be searched by a quantum computer.
This task has several interpretations.
\bit
\item  {\em Computing a Nash equilibrium of a zero-sum game.}   One player's strategies are
specified on a hard drive (the set $X$) and the other have an implicit description (the
set $Y$).  
\item  {\em Linear programming.}  Each $x\in X$ is
a linear constraint and we search over
the set $\Delta(Y)$ for a point satisfying these constraints.
\item {\em Approximate MLE.} Here we let $\Delta(Y)$ be our set of models and we approximate
the sum over $x$ in MLE with the minimum in saddle-point optimization.  This could be
appropriate if the quality of a model were determined by its worst point.
\eit

This problem can also be generalized to the case where $f$ is a concave function
of $y$. In this case, there are a wide number of applications described by
Clarkson~\cite{Clarkson10}.  We discuss a range of additional applications in
\secref{algorithms}.


\section{Known algorithms}\label{sec:known}
The new algorithms in this paper are fairly simple combinations of existing classical
algorithms for data reduction and quantum algorithms for optimization and sampling.  In
this section we will review these algorithms.

\subsection{Quantum optimization algorithms.}\label{sec:known-opt}
 We will consider a few quantum subroutines for
the  problem of maximizing $F(y)$.  These can be seen as quantum analogues of brute-force
search and of simulated annealing.
  \paragraph{Grover/Durr-H\o{}yer~\cite{Grover96,durr96}.} This can find the maximum using $O(\sqrt{m})$
  evaluations of $F$.

    \paragraph{Adiabatic optimization~\cite{farhi00}.} Suppose for simplicity that $m$ is a power of 2, and
  define the Hamiltonians
\ba H(s) &= (1-s)H_0 + s H_F \label{eq:H0HF-def} \\ \nn
 H_0 &= -\sum_{i=1}^{\log(m)} \sigma_x^i,
& H_F &= -\sum_y F(y) \proj y 
.\ea

The algorithm starts with all qubits in the $\ket{+}$ state and evolves under $H(s)$ with
$s$ gradually changing from 0 to 1. (Many variations of this basic idea have also been
proposed including running quickly and/or at variable speeds, starting in different
states, replacing $H_0$, using more complicated paths, running at nonzero temperature
and/or using noisy hardware.)  In general this algorithm can be thought of as a heuristic
since sharp bounds are usually not known on its runtime, or on the minimum spectral gap.
The presence of the $\text{diag}(F)$ term means that running $H(s)$ for time $T$ generally
requires $O(T)$ evaluations of $F$, a point originally made in \cite[Section 5]{farhi00}
and later improved by modern Hamiltonian simulation algorithms such as \cite{H-sim-signal}.

\paragraph{Quantum Approximate Optimization Algorithm (QAOA)~\cite{QAOA,QAOA2}.}  The
  algorithm is parametrized by a positive integer $p$.  The algorithm has a classical
  outer loop which searches over parameters $\theta_1,\ldots,\theta_{2p}$ and an inner
  loop which measures $H_F$ on the state 
\be 
 e^{i\theta_{2p}H_0} e^{i\theta_{2p-1}H_F} 
\cdots e^{i\theta_{2}H_0} e^{i\theta_{1}H_F} \ket{+}^{\ot \log
    m},\ee 
  where $H_0,H_F$ are defined in \cref{eq:H0HF-def}.
In the limit $p\ra \infty$, this includes adiabatic optimization as a special
  case.  Indeed an adiabatic-like schedule could be used as a starting point for the
  search over $\theta$.  The parameter $p$ is analogous to the $T$ in the adiabatic
  algorithm and likewise the inner loop of QAOA uses $O(p)$ evaluations of $F$.

    \paragraph{Quantum Quench~\cite{hastings-quench}.} A variant of the above two algorithms simply
    performs $\exp(i(\alpha H_0 + \beta H_F))$ for some appropriately chosen
    real numbers $\alpha,\beta$.
    
A common theme for each algorithm is that their cost is dominated by evaluations of $F$.
The number of evaluation differs, as does the probability of finding an optimal or
near-optimal choice of $y$.  

\subsection{Sampling algorithms}\label{sec:sampling-review}
We also consider quantum algorithms for the task of sampling from the distribution
$\pi(y) = \exp(F(y))/Z$, given the ability to calculate $F(y)$.  Here
$Z=\sum_y \exp(F(y))$ does not have to be known to the algorithm. We will assume that
$F(y)\leq 0$ for all $y$.  Otherwise we can assume we know an upper bound
$F_{\text{max}}\geq F(y)$ for all $y$ and can replace $F(y)$ with $F(y)-F_{\text{max}}$.

\paragraph{ Rejection sampling.} 
Choose $y$ uniformly at random from $[m]$ and accept with
  probability $\exp(F(y))$, so that the overall acceptance probability is $Z/m$.  On a
  classical computer, this requires on average $m/Z$ repetitions to produce a sample from
  $\pi$, while a quantum computer can produce a sample from $\pi$ using $\sqrt{m/Z}$
  evaluations of $F$.

If we start with $y$ drawn instead from the distribution $\pi_0$ then the acceptance
probability is instead $\frac{\exp(F(y))}{Z'\pi_0(y)}$, where $Z' \geq \max_y
\frac{\exp(F(y))}{\pi_0(y)}$.  The  average acceptance probability is then $Z/Z'$.
This means $O(\sqrt{Z'/Z})$ evaluations of $F$.  Here we need to assume that
$\ket{\sqrt{\pi_0}} = \sum_y \sqrt{\pi_0(y)} \ket{y}$ can be efficiently prepared.
More details can be found in \cite{ozols13}, including 
optimal schemes for approximate or exact rejection sampling.

    \paragraph{Decohering quantum walks~\cite{Richter07a,Richter07b}.} The classical
  Metropolis walk fixes some $d$-regular graph on $Y$ and has transition probability
 $W_{y,y'}=\exp(\min(0, F(y)-F(y')))$ of moving from $y'$ to $y$.  Iterating $W$ on a
 classical computer will
 converge to $\pi$, although with a mixing time that is often hard to characterize.  On a
 quantum computer we can alternate between applying $e^{iWt}$ and measuring.  This
 strategy is conjectured to mix quadratically faster than the classical approach, but this
 has been proven only in special cases.

 \paragraph{Quantum simulated annealing.} For $\beta\in [0,1]$, define the distribution
  $\pi_\beta(y) = \exp(\beta F(y)) / Z_\beta$ where $Z_\beta = \sum_y \exp(\beta F(y))$.
  Suppose that the Metropolis walk on $\pi_\beta$ has spectral gap $\geq g$ for all
  $\beta$.  Then \cite{wocjan08} shows how to sample from $\pi$ in time $\tilde O(g^{-1/2}
  \max_y |F(y)|)$.  This was recently improved in \cite{HW19} to $\tilde O(g^{-1/2}
  \sqrt{\log(Z_0/Z_1)})$.

\subsection{Data reduction (coreset) techniques}\label{sec:coreset-review}
The above quantum algorithms can be directly applied to the first two tasks: optimization
algorithms for MLE and sampling algorithms for Bayesian inference.  (Later we will see
that either type of algorithm also can apply to saddle-point optimization.)  However, in
each case the runtime is dominated by evaluations of $F(y)$ for superpositions of values
$y$.  In general these evaluations require an inner loop that iterates over the entire set
$X$, requiring time $O(n)$. As discussed in \secref{overview}, this reduces the achievable
quantum speedup.  To address this, we will use classical algorithms for data reduction,
which we review here.

\paragraph{Coreset definition.}
Given a data set $X$, a coreset is a pair $(X',w)$ with $X'\subseteq X$ and
$w:X\ra \bbR_{\geq 0}$ a weight function.  We can take $X' := \supp(w)$ so that
$w$ alone is enough to define the coreset, and $|X'| = \|w\|_0$.  
Define $F_{w}$ by \be F_{w}(y) = r(y) +
\sum_{x\in X'} w(x) f(x,y),
\label{eq:F-coreset}\ee
with $r(x),f(x,y)$ defined as above.  We say that $(X',w)$ (or simply $w$)
is an
$\eps$-coreset if 
\be |F(y) - F_{w}(y)| \leq \eps |F(y)|, \quad \forall y\in Y.\ee
For MLE, using an $\eps$-coreset means that the best likelihood based
on the coreset is
within a $1+\eps$ multiplicative factor of the true optimum.  For
Bayesian inference, the probabilities obtained will be within a
multiplicative factor of $e^{\eps|F(y)|}$ of the true
probabilities, which means that highly unlikely events remain fairly
unlikely even with a coreset.  Stronger guarantees on the posterior mean and
variance can also be obtained by using a more application-specific
metric~\cite{HKCB19}. 

We now review several methods for constructing coresets.

\paragraph{Importance sampling.}
A standard approach to constructing coresets is importance sampling.  The idea
here is to estimate the ``importance'' of each element $x\in X$ by an
easy-to-compute function $s(x) \geq 0$.  Let $s_{\text{tot}} = \sum_x s(x)$.  Then
each element in $X'$ is chosen by independently choosing element $x$ with
probability $s(x)/s_{\text{tot}}$.  We repeat this $k$ times, so $|X'|=k$,
possibly taking $X'$ to be a multiset.  To make the resulting estimator
unbiased, we choose $w(x) = \frac{n}{k} \frac{s_{\text{tot}}}{s(x)}$.

This leaves open the question of how we choose the importance weights.  One
approach is by estimating the ``sensitivity'' of each point $x\in X$, defined as
\be \sigma(x) := \max_{y\in Y} \frac{|f(x,y)|}{|F(y)|}.\ee 
While directly
computing even a single $\sigma(x)$ seems to already require a sum over $x$
within a max over $y$, it suffices to use any $s(x)$ satisfying $s(x)\geq
\sigma(x)$, and in this way we can reduce the cost of computing $s(x)$.
However, the coreset size will depend on $s_{\text{tot}}$ and not directly on
the $\sigma(x)$.  To state this result formally we also need to define the
dimension of the query space (following Definition 4.5 of \cite{BFL16} and
building on VC dimension~\cite{VC71}).
\begin{dfn}
  Given $\cX,Y$, $f:\cX\times Y \mapsto \bbR_{\geq 0}$, $y\in Y$ and $r\geq 0$,
  define the level set $L(y,r) = \{ x : f(x,y) \leq r\} \subseteq \cX$.  The {\em dimension}
  of $(\cX,Y,f)$ is the smallest integer $D$ such that for all $S\subseteq \cX$,
  \be \L| \L \{ L(y,r) : y \in Y, r \geq 0 \R \} \R | \leq |S|^D.\ee
\end{dfn}
For $k$-means/medians clustering in $N$-dimensional Euclidean space, we have $D=O(kd)$,
and for general metric spaces with $n$ points we can bound $D=O(k\log
n)$, following the arguments in \cite[Section 6]{BFL16}.

\begin{thm}[Thm 5.5 of \cite{BFL16}]
  Let $\cX,X,Y,f,\sigma,s,s_{\text{tot}},d$ be defined as above and choose a
  coreset $(X',w)$ of size $k$ using the importance sampling procedure above.
  Suppose that
  \be k \geq c\frac{s_{\text{tot}}}{\eps^2} \L( d \log s_{\text{tot}} +
  \log\L(\frac 1 \delta\R)\R),\ee
  with $c>0$ a universal constant.  Then with probability $\geq 1-\delta$, we have
  \be \L|F(y)  - F_w(y)\R| \leq
  \eps \L|F(y)\R| \qquad \forall y\in Y.\ee
\end{thm}
A review of the use of this technique with somewhat weaker bounds but a simpler
exposition can be found in \cite{Bachem17}.

This framework has been used effectively for clustering
problems~\cite{FeldmanL11,Bachem17,BFL16} and for Bayesian inference~\cite{HCB16}.  It remains only to describe how to
compute $s(x)$ efficiently and to bound $s_{\text{tot}}$.  First a fast and
crude approximation is constructed and used to estimate sensitivities.  These
estimates are then used to construct a coreset.  Here the quality of the
approximation determines how close $s(x)$ is to $\sigma(x)$ and thus determines
the size of the coreset, but otherwise does not affect the quality of the final
approximation.  Finally a more careful (perhaps exhaustive) search for
clusterings can be carried out on the coreset.  Details including the resulting
bounds on $s_{\text{tot}}$ are described in more detail in
\Cref{sec:algorithms}.


\paragraph{Adaptive coresets.}
Proposals for adaptive coresets also exist in which $X'$ is built up element by
element~\cite{CB17,CB18,Clarkson10}.  There are many strategies for doing so,
but a general theme is to solve the optimization problem on $X'$ and use the
solution to determine which points from $X$ would be helpful to add next to
$X'$.  One way to choose points for $X'$ is to view $w$ itself as the solution
to an optimization problem, i.e.~we would like to minimize the convex function
\be L_{\text{core}}(w):= \max_y \L| F(y) - F_w(y) \R|, \label{eq:w-objective}\ee subject to
$\|w\|_1=1$ and $\|w\|_0\leq k$ (if we want a coreset of size $\leq k$).

Approximately minimizing convex functions over sparse vectors in the simplex is
a problem that naturally fits the Frank-Wolfe algorithm, also known as the
conditional gradient algorithm~\cite{FW56}. Starting from an initial vector
$w_0$, Frank-Wolfe constructs a series of iterates $w_1,w_2,\ldots$ as follows.
At step $t$, find $x = \arg\min (\grad L_{\text{core}}(w_t))_x$, and then set
$w_{t+1} = (1-\eta_t) w_t + \eta_t e_x$, where $e_x$ is the vector with a 1 in
position $x$ and zeros elsewhere.  In other words, find the best
coordinate direction $x$ in which to move and then take a step towards $x$ of
size $\eta_t>0$.  (This description is for optimizing within the probability
simplex, but the algorithm can be defined for $w$ constrained to any convex
set, as described in \cite{FW56,bubeck15})  Here the $\eta_t$ can be chosen according to a fixed schedule such as
$\eta_t=1/\sqrt t$, by using known properties of $L_{\text{core}}(w)$ such as smoothness and
strong convexity, or adaptively, e.g.~by using a line search.  A key feature is
that the $t^{\text{th}}$ iterate satisfies $\|w\|_0\leq t$, so we can obtain a
coreset of size $k$ by stopping after $k$ iterations.

The problem with the above approach is that, as with computing sensitivities,
the loss function $L_{\text{core}}(w)$ takes as much time to compute as the original problem.
Thus we will instead need to minimize some more efficiently computable proxy for
$L_{\text{core}}(w)$.  Towards this end, interpret $L_{\text{core}}(w)$ as the $\infty$-norm of the vector
\be E_{\text{core}}(w) := \sum_y (F(y) - F_w(y)) e_y \in \bbR^Y.\ee
We will consider two approaches, one suited to Bayesian inference and the other
to saddle-point optimization.

\paragraph{Adaptive Hilbert coresets for Bayesian inference.}
Campbell and Broderick~\cite{CB17,CB18} replace $L_{\text{core}}(w)$
with an appropriately weighted 2-norm of ${E_{\text{core}}(w)}$,
i.e.~they fix an inner product $\langle, \rangle$ on $\bbR^Y$ and seek
to minimize \be L_{\text{CB}}:= \langle E_{\text{core}}(w),
E_{\text{core}}(w)\rangle_{\text{CB}}\ee
One choice of inner product for the problem of Bayesian inference is
\be \langle f,g \rangle_{\pi} := \sum_{y\in Y} \pi(y) f(y)g(y).\ee
Since it is inefficient to evaluate this exactly, we can instead use
$m$ samples $y_1,\ldots,y_m$ from some approximation to $\pi$ and
define
\be \langle f,g\rangle_{\tilde \pi} := m^{-1} \sum_{i\in[m]}
f(y_i)g(y_i).\ee
This allows all computations to be carried out in an $m$-dimensional
space, which can lead to significant savings.  Explicitly we can
define $\Pi(f) = (f(y_1),\ldots,f(y_m))$ so that
$\langle f,g\rangle_{\tilde \pi} = m^{-1} \langle \Pi(f),
\Pi(g)\rangle$.

With this reduced space, it becomes efficient to compute the
Frank-Wolfe updates.  The algorithm needs to keep track of $\Pi(F)$
and $\Pi(F(w))$, which each take space $m$ and can be updated in time
$|X|$ and $|X'|=\|w\|_0$ respectively.  Finding the best $x$ to add to
$X'$ requires computing $\langle F,f(x,\cdot)\rangle_{\tilde \pi}$ and
$\langle F(w),f(x,\cdot)\rangle_{\tilde \pi}$ for each $x\in X$, where
$f(x,\cdot) := \sum_y f(x,y)e_y$.  Each of these inner products takes
time $O(m)$ for a total cost of $O(|X|m)$.  Adding a point to $X'$
also requires choosing its weight but the methods for doing this in
\cite{CB17,CB18} also require time $O(|X|m)$.

  \paragraph{Adaptive coresets from zero-sum games.}
  Grigoriadis and Khachiyan~\cite{GK95} consider a slightly different
  problem in which our goal is instead to minimize
  \be L_{\text{SP}} :=\max_y F_w(y).  \ee
  The notation $L_{\text{SP}}$ is used because this is a minimax or saddle-point
  optimization problem.  Again, direct evaluation of $L_{\text{SP}}$ would be
  too expensive, so we resort to approximations.  First, observe that
  $L_{\text{SP}} = \max_{v\in\Delta(Y)} \langle v,F(w)\rangle$.  Then 
  $\min_w L_{\text{SP}}$ is the minimax problem
  \be \min_{w\in \Delta(X)}\max_{v\in\Delta(Y)}\sum_{x,y} w(x) v(y) f(x,y)\ee
  corresponding to finding an equilibrium of a two-player zero-sum game.  
The idea of \cite{GK95} is then to alternate steps of a stochastic version of
Frank-Wolfe on each of $v$ and $w$.  This has since been recognized as an
example of the multiplicative weights update method~\cite{AHK12} and even more
generally, as
Stochastic Saddle Point Mirror Descent (S-SP-MD)~\cite[Section 6.5]{bubeck15}.
In \secref{algorithms} we will see further applications of these generalizations.

The intuition for the algorithm of \cite{GK95}  comes from viewing $X$ and $Y$ as
strategy sets for a game with payoff $-f(x,y)$ for the $X$ player and $f(x,y)$
for the $Y$ player.  The algorithm samples a series of strategies $x_1,y_1,x_2,y_2, \ldots,
x_t,y_t$,
from distributions $w_1,v_1,\ldots,w_t,v_t$ respectively.  These distributions
are defined in a way that biases them towards responding well to the previous
strategies of the other player.  
\bas
w_t(x)& = \frac{W_t(x)}{\sum_{x'} W_t(x')} \\
W_t(x) & = \exp(-\eta_t(f(x,y_1) + \cdots + f(x,y_{t-1}))) \\
v_t(y) & = \frac{V_t(y)}{\sum_{y'} V_t(y')} \\
V_t(x) & = \exp(\eta_t(f(x_1,y) + \cdots + f(x_t,y))) 
\eas

Here the step size $\eta_t$ can be taken either to be a
fixed $\eps/2$ following \cite{GK95}
or as $\sqrt{2/t}$ following \cite{bubeck15}.  If we set $\eta_t=\eps/2$ then
\cite{GK95} proved that the error will be $\leq \eps$ with probability $\geq
1/2$ after $t$ reaches
$4\max_{x,y} f(x,y)^2 \log(mn)/\eps^2$.  If we set $\eta_t=\sqrt{2/t}$ then a
similar convergence guarantee holds~\cite{bubeck15}, but without needing to fix
in advance the 
number of rounds of the algorithm.

\section{Algorithms}\label{sec:algorithms}
We are now ready to describe our new algorithms and performance guarantees.  Each
algorithm is designed for the hybrid quantum-classical setting described in
\secref{data-models}, addresses one of the tasks from \secref{setting} and uses as subroutines
the quantum and classical algorithms from \secref{known}.  Each
algorithms is described both in full generality and for a specific representative
example.  The goal of this is to show the range of possibilities while also giving a
self-contained presentation of applications.

\medskip\noindent \framebox{
\begin{minipage}{\boxwidth}
{\bf Algorithm 1 (general version)} {\em  Non-adaptive coresets for maximum a posteriori
  estimation.}\\
{\bf Inputs: } Data $X=\{x_1,\ldots,x_n\}$.  This algorithm needs the user to specify a function $f$, a classical
algorithm for generating a coreset (e.g. an approximation algorithm followed by importance
sampling) and a quantum optimization algorithm (e.g. Grover, adiabatic, etc.)\\
\textbf{Output}: $y$, which is likely to be an exact or approximate solution to \eq{MLE}.

\mnb{Algorithm: }
\benum
\item Given input $X$, use the classical algorithm to construct a coreset $(X',w)$.
\item Run the quantum optimization algorithm on $(X',w)$.
\eenum
\end{minipage}}

This ``algorithm'' is more of a framework than a detailed algorithm.  However, it can
readily be adapted to hard and relevant optimization problems, such as the following example.

\medskip\noindent \framebox{
\begin{minipage}{\boxwidth}
{\bf Algorithm 1.1 (specific version)} {\em  Non-adaptive coresets for $k$-means
  clustering.}\\
{\bf Inputs: } Data points $x_1,\ldots,x_n\in \bbR^d$, a number of cluster centers $k$ and
an accuracy parameter $\eps>0$.\\
{\bf Output:} Cluster centers $y_1,\ldots,y_k$ approximately minimizing 
$$\sum_{i\in [n]} \min_{j\in [k]} \|x_i - y_j\|^2.$$

\mnb{Algorithm:}
\benum
\item Use the offline coreset algorithm of \cite{BFL16} to construct a
  coreset $(X',w)$ of size $m = O(\eps^{-2}k\log(k)\min(k/\eps,d))$.
\item Use Grover to search for the best clustering of $(X',w)$.  To make the search space
  finite, we use \cite{IKI94} to reduce the search space to the $O(m^{dk})$
  possible Voronoi partitions.  Given such a partitioning, the cluster center is just the weighted center of those points.
\eenum
\end{minipage}}

The first step of the algorithm takes time $n\cdot \poly(k,d,\eps^{-1})$ time on
the classical computer, while the second step takes the quantum computer time
$m^{dk/2}\cdot \poly(m,k,d)$, since Grover's algorithm requires $O(m^{dk/2})$
iterations and each inner loop requires time $\poly(m,k,d)$.  While classical
computers of course could also make use of the coreset and achieve a runtime of
$(n + m^{kd + O(1)})\cdot \poly(k,d,\eps^{-1})$, the resulting hybrid
classical-quantum algorithm achieves nearly a quadratic speedup over the purely
classical algorithm that also use coresets.

It is important to point out that Algorithm 1.1 is a significant specialization
of Algorithm 1 and that many easy variants apply. For example, we could replace
Grover's algorithm with a heuristic such as the adiabatic algorithm or QAOA.  Or
we could replace the Euclidean $k$-means problem with a generalization known as
M-estimators on metric spaces (see \cite{BFL16}) which can handle more general
geometries, as well as having other properties, such as being robust to
outliers.

Another large class of variants is to use non-adaptive coresets for the other
computational tasks: Bayesian inference and saddle-point optimization.  This
application is fairly straightforward and we will not explore it further in this
paper.

However, there is a sense in which Algorithm 1 in all its flavors does not use
the quantum computer in a very interesting way.  Arguably the algorithm is
mostly classical, with the quantum computer being used only after sophisticated
classical algorithms have reduced the data set to a representative sample.  Our
remaining algorithms will instead use the quantum computer interactively.


\medskip\noindent \framebox{
\begin{minipage}{\boxwidth}
{\bf Algorithm 2 (general version)} {\em  Adaptive coresets for Bayesian inference.}\\
{\bf Inputs: } Data $x_1,\ldots,x_n$.  A description of $\pi_0$, $f$ and a
classical algorithm $\cA$ that takes as input $X$, a coreset $(X',w)$ and a set of samples
$y_1,\ldots,y_k$ and outputs an updated coreset.   A maximum coreset size $m$.  A quantum
algorithm $\cB$ for Bayesian inference.\\
\textbf{Output}: an approximate sample from $\pi_{\text{posterior}}$.

\begin{algorithmic}[1]
\State Initialize $(X',w)$ to be the empty set.
\For{$k\in \{1,\ldots, m\}$}
\State Use the quantum algorithm $\cB$ to sample $y_k$ according to the distribution
\be \pi(y_k) = \exp(F_{X',w}(y_k)) / Z_{X',w},\ee
where $F_{X',w}$ is from \eq{F-coreset} and $Z_{X',w} = \sum_y \exp(F_{X',w}(y))$.
\State Use the classical algorithm $\cA$ to update the coreset $(X',w)$. 
\EndFor
\State Output $y_m$. 
\end{algorithmic}
\end{minipage}}

In many cases the quantum algorithm could also return a q-sample at little or no extra
cost.  A q-sample from a distribution $\pi$ is defined to be the state
\be \sum_y \sqrt{\pi(y)}\ket{y}.\ee
The idea of a q-sample was introduced in \cite{AT-adia-03} and can have significant
advantages over ordinary samples for some applications.  In the examples given in
\secref{sampling-review}, rejection sampling and quantum simulated annealing already
return q-samples while decohering quantum walks do not.

Again we make  Algorithm 2 concrete by specifying $\cA,\cB$ and other features of the
problem.  One representative example is logistic (or logit) regression.  Here we are given a data set
$\{(x_i,\ell_i)\}_{i\in [n]}$ consisting of points $x_i\in\bbR^d$ and
labels $\ell_i\in \{-1,1\}$.   For convenience we assume that the first coordinate of each
$x_i$ is equal to one, leaving $d-1$ effective parameters.
There is a parameter vector $y\in \bbR^d$ that we would like to sample from.  (The reason
sampling from the posterior might be preferable to MLE is that  samples can yield
additional information such as credible intervals.)
 Logistic regression models the probability of a label as
\be p(\ell_i| x_i, y) = \frac{1}{1 + e^{-\ell_i \langle x_i, y\rangle}},\ee
so that $f((x_i,\ell_i), y) = -\log(1 + e^{-\ell_i \langle x_i, y\rangle})$.
For simplicity we can take our prior on $y$ to be $\cN(0,I_d)$, i.e. $d$ independent
Gaussians each with mean 0 and variance 1.  

\medskip\noindent \framebox{
\begin{minipage}{\boxwidth}
{\bf Algorithm 2.1 (specific version)} {\em  Adaptive coresets for logistic regression.}\\
{\bf Inputs: } Data $\{(x_i,\ell_i)\}_{i\in [n]}$\\
\textbf{Output}: An approximate sample from $\pi_{\text{posterior}}$.

\begin{algorithmic}[1]
\State Initialize $(X',w)$ to be the empty set.
\For{$k\in \{1,\ldots, m\}$}
\State Use quantum simulated annealing (see \secref{sampling-review}) to approximately
sample $y_k$ according to the distribution
\be \pi(y_k) = \frac{\exp(F_{X',w}(y_k))}{Z_{X',w}},\ee
where $F_{X',w}$ is from \eq{F-coreset} and $Z_{X',w} = \sum_y \exp(F_{X',w})(y)$.
\State Use the GIGA algorithm (from \secref{coreset-review}) with the 
measure $\hat\pi$ which puts weight $1/k$ at each of the points $y_1,\ldots,y_k$.  This yields a
point to add to $X'$ and a prescription for  updating the weights $w$.  
\EndFor
\State Output $y_m$. 
\end{algorithmic}
\end{minipage}}

The example of logistic regression was taken only for completeness, and we could have
replaced it with any other cheap-to-evaluate model used in machine learning, such as
restricted Boltzmann machines.  In each case, the formal guarantees in \cite{CB17,CB18}
depend on parameters that are often hard to rigorously bound, so this algorithm should be
considered a heuristic.  Empirically testing it on either classical or quantum hardware is
left to future work.

Finally we will see an interactive algorithm with a rigorous performance
guarantee.  This will address the problem of saddle-point optimization, which we
can think of as a zero-sum game with one player's strategy set $X$ described by
a classical database and the other player's strategy set $Y$ accessible in
superposition by a quantum computer.

\medskip\noindent \framebox{
\begin{minipage}{\boxwidth}
{\bf Algorithm 3 (general version)} {\em  Saddle-point optimization \eq{zero-sum}.}\\
{\bf Inputs: } A data set  $X = \{x_1,\ldots,x_n\}$, accuracy parameter $\eps>0$, and code
for a function $f:X\times Y \mapsto [-1,1]$. \\
\textbf{Output}: Distributions $\theta_Y,\theta_X$ approximately achieving the max or min
(respectively) in \eq{zero-sum}.

\begin{algorithmic}[1]
\State Let $T = \lceil \log(|X|\cdot |Y|)/\eps^2 \rceil$.
\For{$t\in \{1,\ldots, T\}$}
\State Use the classical computer to sample $x_t$ according to
\be \Pr[x_t] \propto \exp\left (-\eps\sum_{s< t} f(x_t, y_s)\right).\ee
\State Use a quantum sampling algorithm to sample $y_t$ according to
\be \Pr[y_t] \propto \exp\left (\eps\sum_{s\leq t} f(x_s, y_t)\right).\ee
\EndFor
\State Let $\theta_X$ be the measure that places weight $1/T$ on each of $x_1,\ldots,x_T$. 
\State Let $\theta_Y$ be the measure that places weight $1/T$ on each of $y_1,\ldots,y_T$. 
\end{algorithmic}
\end{minipage}}

According to \cite{GK95}, this algorithm yields an answer within $O(\eps)$ of
the true value of \eq{zero-sum}. To appreciate its efficiency suppose we take
$\eps$ to be constant.  Then the classical algorithm sweeps through the entire
dataset $X$ only $T = O(\log( |X|\cdot |Y|))$ times (with an inner loop of time
$O(T)$), and in any reasonable algorithm it would have to do this at least once.
Likewise the quantum algorithm has an inner loop that takes time only $O(T)$,
and needs to generate only $O(T)$ samples. For both the classical and the
quantum runtime, we are within a factor of $O(T^2)$ of the best possible time we
could expect.  (We note that while algorithms for this problem are known with
nearly $1/\eps$ scaling~\cite{DDK15}, they do not have the sparsity properties
of \cite{GK95} and so cannot be used here.)
In general if we use a Grover-type algorithm for the quantum part (specifically
rejection sampling; cf.~\secref{sampling-review}), the
run-time will be $O(|X|T^2)$ for the classical computer and $O(|Y|^{1/2}T^2)$ for the
quantum computer.  

 For these algorithms to be useful we need
$|Y| \gg |X|$.  Otherwise a classical computer could simulate the quantum
computer in less time than it would take to read the dataset.  In general if
$|Y|=|X|^\beta$ we would expect the speedups plotted in Fig.~\ref{fig:speedup}.

Other recent work~\cite{LCW-training,AG-zero-sum} has also proposed
quantum algorithms for zero-sum games, using a purely oracle data
model.  In this setting, they can also obtain a square-root speedup of
the search over $X$, yielding an overall runtime of $\tilde
O(\sqrt{|X|} + \sqrt{|Y|})$.  These works can be viewed as
specializations of the SDP algorithms of 
\cite{BrandaoS16,1710.02581,1809.00643,1809.01731,vAG18} that take
advantage of the fact that they are working in the probability simplex.

\paragraph{Applications of Algorithm 3.}  While zero-sum games and saddle-point
problems have widespread application (see e.g.~\cite{minimax95}), the challenge
is finding a setting where the data model is natural. (Similar challenges apply
of course to  applying quantum algorithms with other non-standard data models
such as linear systems solvers and the recent quantum algorithms for semidefinite programming and
convex optimization.) Here are several possible
examples that fit our desired data model.
\benum
\item[3.1] {\em Linear programming.} The goal is to optimize over the
  $m$-dimensional probability simplex $Y$ subject to a set of $n$ constraints
  described by $X$.  Each point in $X$ needs to be described by $\ll m$ bits,
  and so should represent a succinct description of an $m$-dimensional vector. 

\item[3.2] {\em $\ell_1$-norm SVM (support vector machine).}  A further
  specialization of the linear programming is a nonlinear SVM.  We are given a
  collection of labeled data points $X$, a map $\varphi$ from $X$ to a feature
  space $Y$, and wish to select a feature vector $f\in \Delta(Y)$ that will
  correctly classify all the points, meaning that
  $\langle \varphi(x),f\rangle \geq 0$.  This is the decision version of the
  problem, and the optimization version instead estimates
  $\theta := \max_f \min_x \langle \varphi(x),f\rangle$ up to additive error
  $\eps$.  If $\theta<0$ then correct classification is impossible and if
  $\theta>0$ then we call $\theta$ the margin of the classifier.  Because
  $f\in\Delta(Y)$ we have $\|f\|_1=1$, and so we call this $\ell_1$ SVM, by
  contrast with the typical case where the feature vector has bounded
  2-norm. The fact that $f\geq 0$ can be addressed by doubling the size of $Y$
  and replacing $\varphi(x)$ with $\varphi(x) \oplus -\varphi(x)$.
Applications of an $\ell_1$-SVM to gene microarray data and
  text classification are described in \cite{GN-SVM-18}, and in
  \cite{Sra-L1SVM} $\ell_1$-SVMs are found to work well on both real
  and synthetic data.
  
\item[3.3] {\em Boosting~\cite{Clarkson10}.}  Suppose that $X$ is a set of
  labeled data and $Y$ is a set of classifiers (or ``decision stumps'').  We
  would like to choose a sparse convex combination of classifiers in $Y$ such
  that no point in $X$ has average error more than $\eps$ higher than the
  optimal combination of classifiers.
  
\item[3.4] {\em Robust optimization~\cite{BBC11}.} Suppose that $X$ is a set of
  possible states of the world and $Y$ is a set of possible strategies.  We
  should think of $X$ as a set on which we do not know the correct probability
  distribution and so we would like to find a mixture of strategies that
  performs well on all elements of $X$.  For example, let $Y$  be a set of
  investment options and let $X$ be
  historical data on asset prices.  The goal is to find a distribution over
  assets whose returns are as large as possible on the entire historical record.
  
\item[3.5] {\em Security games.} Suppose that $X$ is a set of targets and $Y$ is a
  set of defenses.  The defender wants to allocate resources across different
  defense strategies in a way that the weakest target is still well-defended.
  This would make sense in defending a computer network where $X$ is the set of
  ways that an attacker can gain access (different accounts, machines, services,
  etc.) or in preventing credit-card fraud where $X$ is a historical database of
  past known fraudulent transactions.  In this latter case, the minimax model is
  applicable because any profitable attack can be scaled up.  More
  speculatively,  the goal could be disease eradication (e.g.~for polio), $X$
   a set of known disease locations and $Y$ a set of eradication strategies.  In
   each case, the solution corresponds to actual resources that are allocated
   and so it makes sense that this solution should be sparse.
   
 \item[3.6] {\em Approximate Carath\'{e}odory in $\ell_\infty$
     norm~\cite{Clarkson10}.}  Let $X$ be a finite subset of $\bbR^m$.
   Carath\'{e}odory's theorem states that any vector
   

\eenum

It is tempting to attempt to apply Algorithm 3 to semidefinite
programming where $Y$ is the space of trace-1 psd matrices and $X$ is
a collection of succinctly described linear constraints.  However, the
algorithms used for this
problem~\cite{BrandaoS16,1710.02581,1809.00643,1809.01731,vAG18} all
require some form of Grover search over $X$ for a constraint that is
violated by a given state $\rho$.  This does not appear compatible
with an input model where $X$ is a classical dataset.
By using oracle data models, the various quantum SDP solvers all work equally
well for LPs and SDPs, but when $X$ is classical, LPs appear to be
easier than SDPs.


For each algorithm the correctness guarantees follow essentially immediately
from those of the various subroutines.  In some cases, we may use heuristic
algorithms for some of the subroutines, such as adiabatic optimization or the
GIGA algorithm for coresets.  In that case, the final algorithm would also be
essentially a heuristic.  It would have clear bounds on runtime (e.g.~because
the number of points in the coreset is chosen by the user) but without provable
accuracy guarantees.

\section{Comparison with other algorithms}\label{sec:compare}

How do the new algorithms presented in this paper compare with
previous approaches to these problems, and to previous hybrid
quantum-classical algorithms?

\subsection{Comparison with variational
  algorithms}\label{sec:compare-VQE}
As discussed at the end of \Cref{sec:data-models}, variational
algorithms use a classical outer loop to perform gradient descent on
the circuit parameters of a quantum inner loop.  These can be
extremely general and in some cases amount to performing a local
search over the set of all short circuits implementable in a
particular hardware model.  As a result, they often lack the provable
guarantees of algorithms in this paper.  On the other hand, they can
be run on even very simple quantum computers and running them can
teach us about what we might expect from future quantum hardware.  

While existing variational algorithms are not designed for a setting
with a large classical dataset, there are some connections with the
algorithms in this paper.
Our algorithm 3 can be thought of as running the Frank-Wolfe variant of mirror descent
on the weight vector of the coreset.  This makes the data-reduction approach closer to
the variational algorithms that have recently become nearly synonymous with NISQ
algorithms.  However, there are two key features of our Algorithm 3 that are not suggested
by the usual variational formulation: 1) the ansatz and the classical outer loop
are structured carefully to present the quantum computer with a very limited subset of the
overall data set, and; 2) the output of the quantum computer is usable for a form of
stochastic mirror descent without any of the dimension dependence that is seen in general (e.g.~~\cite{HN19}).

\subsection{Comparison with stochastic gradient descent}
 Many of the problems with handling large datasets
are also relevant to classical computers.  For this reason, when training
continuously parametrized models on large datasets, the standard classical
algorithm is not gradient descent but stochastic gradient descent (SGD).  One
common version of SGD is to sample a single data point at a time and take a
gradient step based on that point.  We might imagine using this for the inner
loop if the outer loop were, say, the Durr-H\o{}yer minimization
algorithm.  One can also interpret between this form of SGD and the usual gradient
descent with mini-batch gradient descent, which samples a set of $k
\ll n = |X|$ points at a time and uses these for gradient estimates.
We use the term ``SGD'' to refer to this last approach.

Mini-batch gradient descent is most directly comparable to using a
coreset of size $k$.  Traditionally, coresets are sampled using some
form of importance sampling while mini-batches are often uniformly
sampled. However, this is not necessary, and there have been proposals
to use importance sampling also to construct
mini-batches~\cite{CRminibatch18}.

The essential difference between SGD and coresets is that coresets are
sampled once and then used throughout the optimization, while SGD
draws fresh samples for each gradient step.  For classical gradient
descent, this difference may not be important, or it may favor
SGD. However, when used as a subroutine inside a Grover or
Durr-H\o{}yer search, the stochastic noise introduced by SGD can be
harmful.
Indeed, the Grover speedup is known to vanish when the oracle is
stochastic and has a non-negligible chance of being replaced by the
identity operator~\cite{RS08}.  To see this, we briefly review an
argument from \cite{RS08}.   Normally Grover
search consists of a series of alternating reflections
$R_SR_OR_SR_OR_SR_O\cdots = (R_SR_O)^T$, where $R_S$ reflects about the starting
state and $R_O$ is the oracle call.  If a single $R_O$ term is
deleted, say in the $j^{\text{th}}$ position, then this sequence
becomes equivalent to $(R_SR_O)^j (R_OR_S)^{T-j}$, using $R_O^2=I$.
In other words, a single deletion reverses the order of all later
rotations.  If each $R_SR_O$ rotates by an angle $\theta \sim 1/T$ then
$R_OR_S$ rotates by $-\theta$ and random deletions result in
alternating between these two.  Overall this random walk will take
time $\sim T^2$ to rotate by an $\Omega(1)$ angle, thus negating the
Grover speedup.  If the failure rate is $\eps \ll 1$ then we can view
this as a random walk with step size $1/T\eps$, so the total time
becomes $T^2\eps$ (or $T$, whichever is greater).
Of course one may consider other strategies, but
\cite{RS08} prove that nothing asymptotically better is possible.
This restriction applies only to stochastic oracles.  If the oracle
produces a coherent superposition of success and failure then the
Grover speedup is possible~\cite{BHMT02}, but in our setting that
would require superposition access to the entire dataset.

Returning now to SGD vs coresets, we can see how SGD is equivalent to
a fault Grover oracle by considering a simple toy model.  Suppose we
would like to estimate $\max_{y\in[m]} F(y)$ with $F(y)=
\sum_{x\in[n]} f(x,y)$ and we are promised that each $f(x,y)\in\{0,1\}$
and that
\be
F(y) = \begin{cases} n\epsilon & y = y_* \\
    0 & \text{otherwise}\end{cases}.
\ee
In other words, there is a subset of $n\eps$ values of $x$ for which
$f(x,y_*)=1$.  For all other values of $x,y$, we have $f=0$.

If we use a coreset, then a set of size $1/\eps$ will have constant
probability of hitting a good value of $x$.  We can then find $y_*$
with constant probability in
time $\sim\sqrt{m}/\eps$.

Using SGD with a batch size of $1/\eps$ results in an estimator for
$F(y)$ with constant probability of success.  This would result in
time $\sim m/\eps$ to find $y_*$, which would be no better than in the
classical case.  If we were to stop early, say after $\sqrt{m}/\eps$
iterations, then Grover would output a nearly uniformly random value
of $y$, yielding essentially no information.
The only way to achieve the SGD speedup would be to
increase the batch size to the point where the oracle had negligible
probability of failure, i.e. to size $\log(m)/\eps$. This would result
in a runtime of $\sqrt{m}\log(m)/\eps$.

This is an example of a more general problem.  A subset of size $k$,
whether used as a coreset or a SGD minibatch, may be ``bad'' with
probability $\delta$.  For coreset algorithms this means the overall
algorithm fails with probability $\delta$.  For a Grover search using
SGD, this means the overall algorithm fails with probability $\sim
\sqrt{|Y|}\delta$.  (Other algorithms for searching over $Y$, such as
the adiabatic algorithm, may have a more complicated dependence on
stochastic noise, and studying this difference is an important open
question.)  Coresets also fail more gracefully. In more complicated
problems where the true max is not only 0 or 1 (as in our example), we
can find the true maximum of $F_w$ even for a nonoptimal choice of
$w$, while Grover with an SGD inner loop would return almost no
information if run for too little time.

In classical optimization algorithms, SGD can also be improved by
variance reduction techniques, such as
\cite{JZ-variance-reduction}. However, these techniques do not seem
obviously compatible with the hybrid approach in this paper.  For
example, \cite{JZ-variance-reduction} needs to occasionally calculate
a full gradient, which in our terminology would take time $O(mn)$.

Recently there has been enormous progress in our theoretical and
practical understanding of more sophisticated variants of gradient
descent~\cite{Ruder-grad}, and it is an important open question to
understand what potential these have for benefiting quantum algorithms.

\section{Conclusion}\label{sec:concl}

This paper has a simple message.  When quantum black-box optimization/sampling
algorithms rely on computing a function in the inner loop, and that function
involves a classical data set, we can use classical data-reduction techniques to
reduce the effective size of this set.  Without doing so, we could expect many
quantum speedups to be significantly weakened, and Grover-type speedups would
become effectively useless.

A more optimistic lesson from the paper is about how to design quantum
algorithms for machine learning tasks.  If we were to be as generous as possible
to the power of quantum computers, we might hope that BQP=PSPACE.  Even in this
case, we would still have the problem that near-term quantum computers will have
a small number of qubits (say $n$) and because of decoherence will not able to
run many gates (say $T$).  (Note that even using FTQC, there is a $\poly\log T$
overhead in the number of qubits, which will still present an effective upper
limit on the number of gates.)  Thus such a quantum computer could only handle
$O(T)$ pieces of classical data.  Without good data-reduction techniques,
fitting a model to a large data set would not obviously be sped up by a small
quantum computer, even with the assumption that BQP=PSPACE.  Conversely, this
optimistic assumption on the power of quantum computers could be a good place to
look for new algorithms, since if quantum computers cannot help in this model,
they will never be useful. The current paper could be seen as an answer to the
question of ``suppose that the adiabatic algorithm worked as well as we could
hope in every case; then how would we use it on practical problems?''

This work opens up many possible open problems.  First, it would be useful to
empirically test some of the many algorithms proposed here.  While many coreset
algorithms have been tested classically, they have generally not be tested in
the regimes where quantum advantage would be expected.

Second, there are more general forms of data reduction that could be
explored. One direction that is orthogonal to coresets would be to reduce the
dimension of the data points.  Here too it seems likely that adaptive schemes
could improve on non-interactive algorithms.

Finally there are likely to be many more ways of developing hybrid classical-quantum
algorithms with nontrivial interaction between the two devices.   The field of quantum
simulation already has many proposed quantum algorithms as well as many classical
algorithms but there has been little research to date on useful mergers of the two.  One
promising approach in this direction is \cite{RWCRPS18} which speeds up Hamiltonian
simulation by taking suitably weighted samples of the terms, although
achieving this with a realistic input data model is still an open question.

\section*{Acknowledgments}
I am grateful for helpful conversations with Olivier Bachem, Trevor
Campbell, Tongyang
Li, and Ben Recht.  My
funding is from NSF grants CCF-1452616, CCF-1729369, PHY-1818914 and ARO
contract W911NF-17-1-0433 and the MIT-IBM Watson AI Lab under the project {\it
  Machine Learning in Hilbert space}. This work was partially done while hosted
at the Institut Henri Poincar\'e and the Institut des Hautes \'Etudes
Scientifiques (while supported by a CARMIN fellowship) and by University College
London.

\end{document}